\documentclass[%
 aip,
 amsmath,amssymb,
 reprint,%
]{revtex4-1}

\usepackage{graphicx}
\usepackage{dcolumn}
\usepackage{bm}

\usepackage[utf8]{inputenc}
\usepackage[T1]{fontenc}
\usepackage{mathptmx}
\usepackage{etoolbox}

\makeatletter
\def\@email#1#2{%
 \endgroup
 \patchcmd{\titleblock@produce}
  {\frontmatter@RRAPformat}
  {\frontmatter@RRAPformat{\produce@RRAP{*#1\href{mailto:#2}{#2}}}\frontmatter@RRAPformat}
  {}{}
}%
\makeatother
\begin{document}

\preprint{AIP/123-QED}

\title{A gate- and flux-controlled supercurrent diode}

\author{F. Paolucci}%
\email{federico.paolucci@pi.infn.it}
\affiliation{ 
INFN Sezione di Pisa, Largo Bruno Pontecorvo 3, I-56127, Pisa, Italy
}%
\affiliation{ 
NEST, Istituto Nanoscienze-CNR and Scuola Normale Superiore, I-56127 Pisa, Italy
}
\author{G. De Simoni}%
\affiliation{ 
NEST, Istituto Nanoscienze-CNR and Scuola Normale Superiore, I-56127 Pisa, Italy
}%

\author{F. Giazotto}%
\email{francesco.giazotto@sns.it}
\affiliation{ 
NEST, Istituto Nanoscienze-CNR and Scuola Normale Superiore, I-56127 Pisa, Italy
}%


\begin{abstract}

Non-reciprocal charge transport in supercurrent diodes (SDs) polarized growing interest in the last few years for its potential applications in superconducting electronics (SCE). So far, SD effects have been reported in complex hybrid superconductor/semiconductor structures or metallic systems subject to moderate magnetic fields, thus showing a limited potentiality for practical applications in SCE. Here, we report the design and the realization of a monolithic SD by exploiting a Dayem bridge-based superconducting quantum interference device (SQUID). Our structure allows reaching rectification efficiencies ($\eta$) up to $\sim6\%$. Moreover, the absolute value and the polarity of $\eta$ can be selected on demand by the modulation of an external magnetic flux or by a gate voltage, thereby guaranteeing high versatility and improved switching speed. Furthermore, our SD operates in a wide range of temperatures up to about the $70\%$ of the superconducting critical temperature of the titanium film composing the interferometer. Our SD can find extended applications in SCE by operating in synergy with widespread superconducting technologies, such as nanocryotrons, rapid single flux quanta (RSFQs) and memories.
\end{abstract}

\maketitle

\begin{figure}
\centering
\includegraphics[width=0.95\columnwidth]{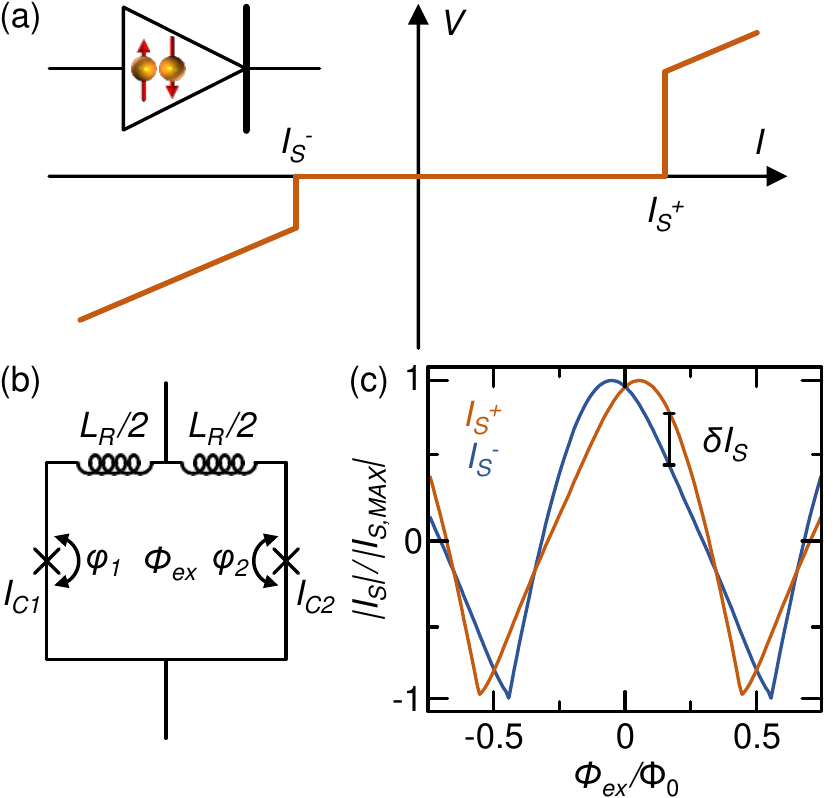}
\caption{\label{Fig1} (a) Schematic voltage ($V$) versus current ($I$) characteristic of a supercurrent diode (SD): the positive ($I_S^+$) and negative ($I_S^-$) switching currents are different. Inset: distinctive shape of a SD. (b) Illustration of a dc-SQUID implementing a SD, where the superconducting ring of inductance $L_R$ is interrupted by two junctions of critical current $I_{C1}$ and $I_{C2}$. $\Phi_{ex}$ is the external magnetic flux, while $\varphi_1$ and $\varphi_1$ denote the phase drops across the left and right junction, respectively. (c) Absolute value of $I_S^+$ and $I_S^-$ for the dc-SQUID calculated for $\beta = 6$ and $I_{C1}=0.5I_{C2}$. $\delta I_S=|I_S^+|-|I_S^-|$ is shown.}
\end{figure}

Superconducting electronics (SCE) aspires to substitute the semiconductor technology thanks to its superior operation speed and improved energy efficiency \cite{Duzer1999,Braginski2019}.
To this end, several superconducting counterparts of the widespread semiconductor electronic devices have been developed. 
Indeed, supercurrent transistors have been realized in the form of cryotrons \cite{Buck1956}, nanocryotrons \cite{McCaughan2014}, rapid single flux quanta (RSFQs) \cite{Likharev1991}, superconducting field-effect transistors (SuFETs) \cite{Nishino1989}, Josephson field-effect transistors (JoFETs) \cite{Clark1980} and fully metallic gated devices\cite{DeSimoni2018,Paolucci2019,Ritter2021}.  
Furthermore, information can be stored by superconducting memories exploiting kinetic inductance \cite{Chen1992}, phase-slips in Josephson junctions \cite{Ligato2021}, superconducting quantum interference devices (SQUIDs) \cite{Murphy2017} and superconductor/ferromagnet hybrid structures \cite{Vernik2013}. 
By contrast, devices implementing non-reciprocal Cooper pairs transport are still at their infancy. Indeed, the first supercurrent diode (SD), i.e. a device showing different positive ($I_S^+$) and negative ($I_S^-$) values of the switching current [see Fig. \ref{Fig1}(a)], has been only recently demonstrated in artificial superconducting superlattices \cite{Ando2020,Miyasaka2021}. Other realizations rely on proximitized two-dimensional electron systems \cite{Baumgartner2022,Baumgartner202b2,Pal2021,Gupta2022,Turini2022}, van der Waals heterostructures \cite{Wu2022,Shin2021,Lin2021}, superconducting transition metal dichalcogenides \cite{Bauriedl2021} and magic angle twisted bilayer graphene \cite{Merida2021}. All these systems showed limited supercurrent rectification efficiencies [$\eta=(|I_S^+|-|I_S^-|)/(|I_S^+|+|I_S^-|)$] reaching values $\eta\sim1\%$.
Despite the great interest in basic science, the exploitation of these exotic systems in SCE practical architectures seems, however, hardly plausible.
Differently, the SD effect shown in conventional superconductor thin films immersed in moderate magnetic fields \cite{Hou2022} suffers from flux trapping, and slow tunability of the rectification absolute value and polarity.

In this Letter, we propose and experimentally demonstrate how to realize a SD by means of conventional metallic superconductors. To this end, we exploit a SQUID interferometer \cite{Clarke2004,Duzer1999} with a large ring inductance ($L_R$), as schematically shown in Fig. \ref{Fig1}(b). As a consequence, the charge transport of our SD can be described through the  resistively-shunted junction (RSJ) \cite{Barone1982}
model
\begin{equation}
   \begin{cases}
    \frac{I_{pass}}{2}+I_{circ}=I_1\left(\varphi_1\right)+\frac{\Phi_0}{2\pi R_1}\dot{\varphi_1}      \\
    \frac{I_{pass}}{2}-I_{circ}=I_2\left(\varphi_2\right)+\frac{\Phi_0}{2\pi R_2}\dot{\varphi_2},
  \end{cases}
    \label{RSJ}
\end{equation}
where $I_{pass}$ is the supercurrent passing through the SQUID, $I_{circ}$ is the current circulating in the ring, $\Phi_0=2.0678\times 10^{-15}$ Wb is the magnetic flux quantum, while $R_i$ and $\varphi_i$ (with $i=1,2$) are the normal-state resistance and the phase drop across the $i$-th junction, respectively. In the simplest case, the two Josephson junctions show a sinusoidal current-to-phase relation (CPR) \cite{Golubov2004}, that is $I_{i}(\varphi_i)=I_{Ci}\sin{\left(\varphi_i\right)}$ (with $i=1,2$), where $I_{Ci}$ is the critical current of the $i$-th junction. The phase drops across the two junctions are connected through the fluxoid quantization in the superconducting ring by
\begin{equation}
 \varphi_2-\varphi_1=2\pi\frac{\Phi_{ex}}{\Phi_0}+\beta I_{circ}+2\pi k,
 \label{fluxquant}
\end{equation}
where $\Phi_{ex}$ is the external magnetic flux, and $k$ is an integer number. The screening parameter accounts for the finite $L_R$ and is defined as $\beta=\frac{2\pi L_R \bar{I_{C}}}{\Phi_0}$, where $\bar{I_{C}}\simeq(I_{C1}+I_{C2})/2$ is the average value of the critical current of the Josephson junctions.

\begin{figure}
\centering
\includegraphics[width=0.95\columnwidth]{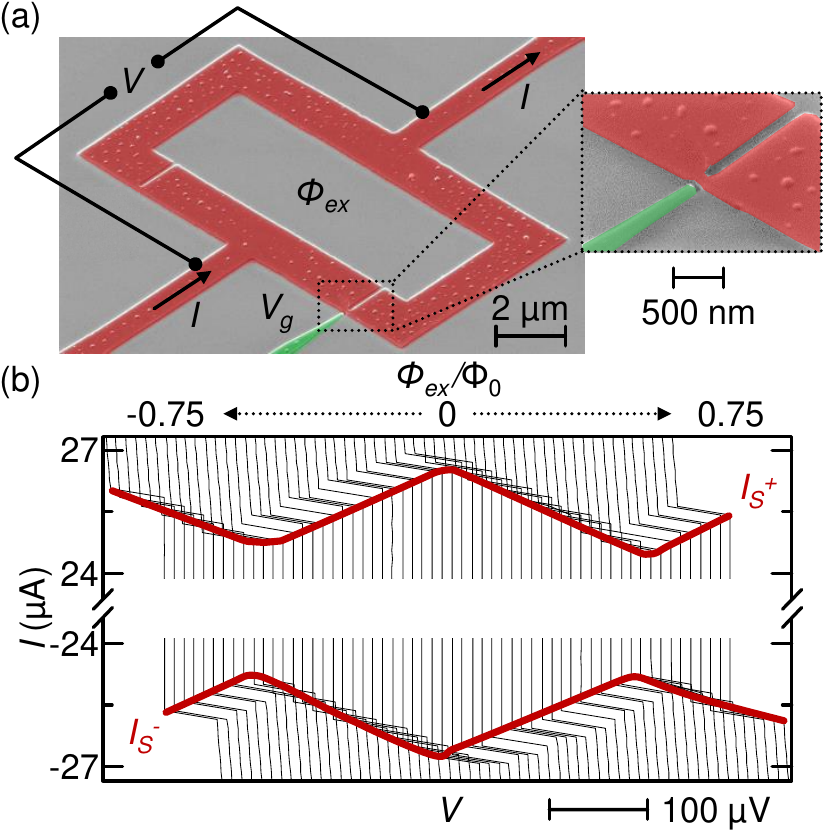}
\caption{\label{Fig2} (a) False-color scanning electron micrograph of a magnetic flux ($\Phi_{ex}$) and gate ($V_g$) controlled SD. The SD is current biased ($I$), while the voltage ($V$) is measured in a 4-terminal configuration. Inset: blow-up of the gated Josephson junction. (b) Back and forth $I$ versus $V$ recorded for different values of $\Phi_{ex}$ at $T=50$ mK. The curves are horizontally offset for clarity. The red line is the fit of $I_S^+$ and $I_S^-$ obtained through the RSJ \cite{Barone1982} and KO-1  \cite{Golubov2004,Kulik1975} models.}
\end{figure}

At a given value of $\Phi_{ex}$, $I_S^+$ ($I_S^-$) can be calculated by maximizing (minimizing) $I_{pass}$ with respect to $\varphi_1$. The combination of an asymmetry of the junctions critical current ($\delta I_C=I_{C1}-I_{C2}\neq0$) and a finite screening parameter $\beta$ causes a shift $\delta\Phi=\frac{\beta\Phi_0}{\pi}\frac{\delta I_C}{\bar{I_{C}}}$ of the $|I_S^{+}|$ and $|I_S^{-}|$ maxima in the opposite direction on the $\Phi_{ex}$ axis. As a consequence, the positive and negative branches of the switching current are different [$\delta I_S(\Phi_{ex})=|I_S^+(\Phi_{ex})|-|I_S^-(\Phi_{ex})|\neq 0$], and the device operates as a SD, as shown in Fig. \ref{Fig1}(c) for $I_{C1}=0.5I_{C2}$ and $\beta=6$. We stress that $\delta I_S$ is positive or negative depending on the half-period of the $I_S(\Phi_{ex})$ interference pattern. 

The SQUID-based geometry allows to design SDs able to tune the rectification through an external magnetic flux, and the asymmetry in the critical current of the Josephson junctions embedded in the ring. The latter could be controlled by employing gated superconductor/normal conductor/superconductor (SNS) Josephson junctions with a low charge carrier material as normal conductor, such as graphene \cite{Girit2009} or a semiconductor nanowire \cite{Spathis2011}. This solution would need hetero-junctions between different materials with non-scalable fabrication protocols. This limitation can be easily bypassed by exploiting the gate-induced suppression of the supercurrent in conventional  \textit{metallic} superconductors \cite{DeSimoni2018,Paolucci2019,Ritter2021}. This effect has been also demonstrated in nano-constriction Josephson junctions (Dayem bridges DBs) \cite{Paolucci2018,Paolucci2019c,DeSimoni2020} and monolithic interferometers \cite{Paolucci2019b}. Despite the microscopic mechanism at the origin of the above gating effect is still 
under debate \cite{DeSimoni2018,Paolucci2019,Ritter2021,Alegria2021,Basset2021,Kafanov2021,Elaily2021}, 
it can be exploited to tune the unbalance between $I_{C1}$ and $I_{C2}$. In particular, we will focus on SDs constituted by gated DB-based SQUIDs. 

Figure \ref{Fig2}(a) shows a false color scanning electron micrograph of a typical flux- and gate-tunable SD. The devices were fabricated by a single step electron beam lithography (EBL) followed by the evaporation of a titanium thin film (thickness $\sim30$ nm and critical temperature $T_{C,Ti}\simeq420$ mK) in the ultra-high vacuum chamber (base pressure $\sim10^{-11}$ torr) of an electron beam evaporator. The superconducting ring (red) is interrupted by two DB Josephson junctions (length and width $\sim150$ nm). The critical current $I_{C1}$ of DB$_1$ can be controlled by a voltage ($V_g$) applied to a local gate electrode (green) placed at a distance of about 30 nm. The current-versus-voltage ($IV$) characteristics of the SD were recorded in a 4-wire configuration in a filtered He$^3$-He$^4$ dilution refrigerator. Figure \ref{Fig2}(b) shows the modulation of the device $IV$ characteristics with $\Phi_{ex}$ recorded at a temperature $T=50$ mK. The triangular magnetic flux pattern of $I_S^+$ ($I_S^-$) is shifted on the magnetic flux axis of a quantity $\delta\Phi_{ex}^+\simeq5\times10^{-2}\Phi_0$ ($\delta\Phi_{ex}^-\simeq-5\times10^{-2}\Phi_0$), thus indicating 
a large $L_R$ value and a 
sizeable asymmetry between the critical currents of the two junctions. Indeed, we extracted $\beta\sim30$ and $I_{C1}\simeq1.03I_{C2}$ from a fit with the RSJ model of Eq. \ref{RSJ}, where the DB Josephson junctions are described by the zero-temperature Kulik-Omel'Yanchuck CPR for a diffusive short junction (KO-1) \cite{Golubov2004,Kulik1975} $I_{i,KO}=I_{Ci}\cos{\left(\frac{\varphi_i}{2}\right)}\operatorname{artanh}\big[{\sin{\left(\frac{\varphi_i}{2}\right)}}\big]$ (with $i=1,2$).

\begin{figure*}
\includegraphics{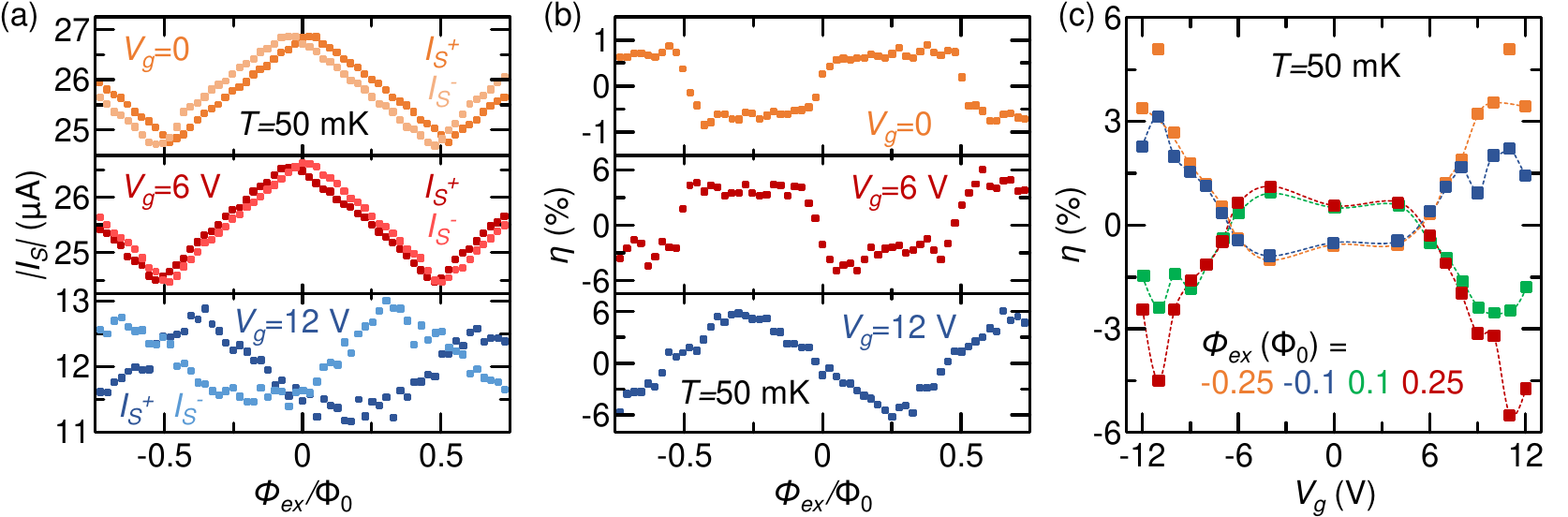}
\caption{\label{Fig3} (a) Absolute value of the positive ($I_S^+$) and negative ($I_S^-$) switching current versus the external magnetic flux ($\Phi_{ex}$) recorded at $T=50$ mK for $V_g=0$ (orange), $V_g=6$ V (red) and $V_g=12$ V (blue). (b) Rectification efficiency ($\eta$) versus the external magnetic flux ($\Phi_{ex}$) extracted for the data in (a). (c) Rectification efficiency ($\eta$) versus gate voltage ($V_g$) obtained for different values of $\Phi_{ex}$ at $T=50$ mK. The dashed lines are guides for the eye.}
\end{figure*}

In order to quantify the rectification performance of our SD, we analyze the switching current interference patterns of our device by varying the voltage applied to the gate electrode.
Figure \ref{Fig3}(a) shows the absolute value of $|I_S^{+}|$ and $|I_S^{-}|$ of the SD recorded at $T=50$ mK for selected values of $V_g$. The pristine device ($V_g=0$) exhibits $\delta I_S > 0$ ($\delta I_S < 0$) in the flux range $0<\Phi_{ex}<0.5\Phi_0$ ($-0.5\Phi_0<\Phi_{ex}<0$). 
The application of a gate voltage at DB$_1$ results in the decrease of $I_{C1}$, thus the absolute value of both polarities of the switching current decreases. Furthermore, the maximum of  $|I_S^+|$ ($|I_S^-|$) moves towards negative (positive) values of $\Phi_{ex}$.
As a consequence, the positive (negative) branch of the switching current dominates in the negative (positive) $\Phi_{ex}$ half-period. 
This behavior is highlighted by the most relevant figure of merit of a SD: the supercurrent rectification efficiency [$\eta=\delta I_S/(|I_S^+|+|I_S^-|)$]. Indeed, $\eta$ changes its polarity by increasing $V_g$, as shown in Fig. \ref{Fig3}(b). Furthermore, the $\eta (\Phi_{ex})$ characteristics transforms from a square wave at low values of gate voltage to a triangular wave at $V_g\geq10$ V, due to the change of the $I_S^{+,-}(\Phi_{ex})$ interference patterns. 

Figure \ref{Fig3}(c) shows the dependence of the rectification efficiency on the gate voltage for selected values of $\Phi_{ex}$. For all values of the external magnetic flux, the maximum value of both polarities of the rectification efficiency rises by increasing gate voltage until reaching its maximum value $\eta\simeq\pm6\%$ for $V_g=11$ V at $\Phi_{ex}=\mp\Phi_0/4$. The rectification efficiency is almost symmetric in the polarity of $V_g$, since the critical current suppression in a metallic superconductor does not depend on the sign of the gate voltage, i.e., it is \textit{bipolar} in the gate voltage \cite{DeSimoni2018,Paolucci2019,Ritter2021}.
These rectification values are several times larger than those obtained in other exotic superconducting systems \cite{Ando2020,Miyasaka2021,Baumgartner2022,Baumgartner202b2,Pal2021,Gupta2022,Turini2022,Wu2022,Shin2021,Lin2021,Bauriedl2021,Merida2021} and smaller than the values reported for metallic superconductors immersed in moderate magnetic fields \cite{Hou2022}. The main advantage of our SD is the possibility to both tune the value of $\eta$ and invert its polarity through $\Phi_{ex}$ and $V_g$. This opportunity enables the possibility to exploit  our device in environments where appreciable magnetic fields are detrimental. Furthermore, our SD guarantees, in principle, high operation speed thank to the fast gate control (as compared to the slower magnetic field control). 

The temperature strongly affects the SD performance, since the critical currents of the two junctions ($I_{C1}$ and $I_{C2}$) are suppressed by the temperature. On the one hand, this implies that the value of $\beta$ lowers by increasing the temperature \cite{Barone1982}, thus the shift of the $I_S$ interference along the $\Phi_{ex}$ axis  decreases \cite{Paolucci2019b}. As a consequence, the rectification efficiency of the SD is expected to reduce ($\eta\to0$ for $\delta\Phi_{ex}\to0$). 
On the other hand, the critical currents ($I_{C1}$ and $I_{C2}$) of two junctions can show a slightly different temperature dependence. In particular, $\delta I_C$ can rise with temperature, resulting in the increase of the shift of the interference pattern along the $\Phi_{ex}$ axis ($\delta\Phi \propto \delta I_C$).
As a consequence, both $\delta I_S$ and $\eta$ could increase with $T$ in a specific temperature range.
Therefore, the rectification efficiency could have a non-trivial dependence on temperature and gate voltage. 

\begin{figure}
\centering
\includegraphics[width=0.95\columnwidth]{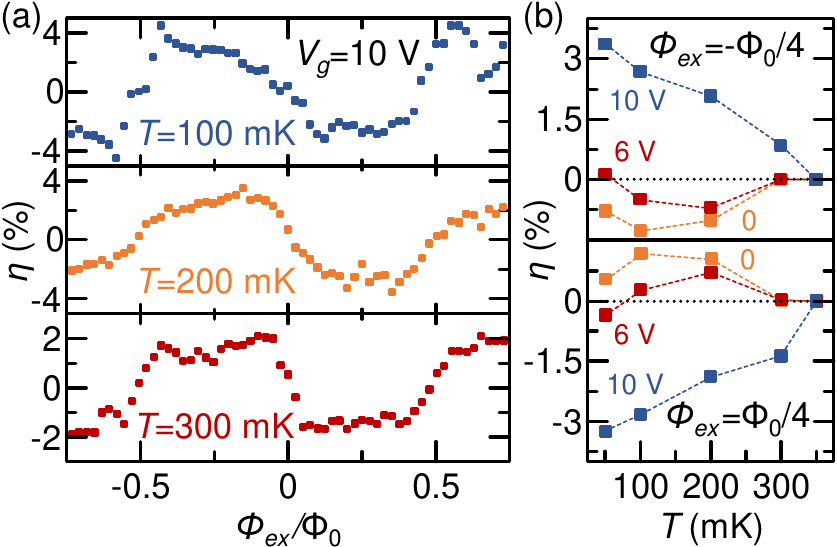}
\caption{\label{Fig4} (a) Rectification efficiency ($\eta$) versus external magnetic flux ($\Phi_{ex}$) for $V_g=10$ V at $T=100$ mK (blue), $T=200$ mK (orange) and $T=300$ mK (red). (b) Temperature ($T$) dependence of the rectification efficiency ($\eta$) recorded at $\Phi_{ex}=-\Phi_0/4$ (top) and $\Phi_{ex}=\Phi_0/4$ (bottom) for selected values of the gate voltage ($V_g$). The dashed lines are guides for the eye.}
\end{figure}

In order to study the full temperature behaviour of our SD, we measured its $I_S^{+,-} (\Phi_{ex})$ characteristics as a function of $V_g$ at values of bath temperature ranging from 50 mK to 350 mK. Figure \ref{Fig4}(a) shows the dependence of $\eta$ on $\Phi_{ex}$ recorded at $V_g=10$ V for $T=100$ mK (blue), $T=200$ mK (orange) and $T=300$ mK (red). On the one hand, the polarity of the rectification of the SD does not change with temperature. On the other hand, the $\eta(\Phi_{ex})$ characteristics changes from a \textit{triangular} to a \textit{square} wave by increasing the temperature, since the decrease of $\beta$ entails a $I_S(\Phi_{ex})$ behavior similar to that measured for low values of $V_g$ at the base temperature. As a consequence, the maximum value of the rectification of our SD at $V_g=10$ V decreases with increasing temperature. This behavior is highlighted in Fig. \ref{Fig4}(b), where $\eta$ is plotted versus $T$ for selected values of $V_g$ at $\Phi_{ex}=-\Phi_0/4$ (top) and $\Phi_{ex}=\Phi_0/4$ (bottom). Indeed, $\eta(V_g=10 \textrm{ V})$ monotonically decreases with temperature, until its full suppression at $T=350$ mK.
Interestingly, the behavior of the SD for $V_g=6$ V is qualitatively different. Specifically, the SD rectification changes its polarity with increasing temperature, since the flux shift due to the application of $V_g$ becomes negligible. Instead, $\eta$ does not change sign for $V_g=0$, but it shows a non-monotonic temperature dependence suggesting that $\delta I_C$ raises by increasing $T$.
Finally, we stress that the SD efficiently works up to $T\simeq300$ mK, i.e., about $\sim 70\%$ of the critical temperature of the titanium thin film. 

In summary, we have designed and realized an original type of supercurrent diode controlled  via an external magnetic flux, and a voltage applied to a gate electrode capacitively-coupled to one of the junctions of the interferometer. To realize such a SD, we exploit two main ingredients: (i) a SQUID with a large screening parameter $\beta$ \cite{Barone1982,Clarke2004}, and asymmetric Dayem bridge Josephson junctions \cite{Golubov2004}, (ii)  gate control of the supercurrent in metallic superconductors\cite{DeSimoni2018,Paolucci2019,Ritter2021}. Our SD shows rectification efficiencies reaching about $6\%$ for both supercurrent directions with the possibility to change the polarity of $\eta$ by both $\Phi_{ex}$ and $V_g$.
Moreover, we have demonstrated supercurrent rectification up to $T=300$ mK, that is around $70\%$ of the critical temperature of the superconductor composing the device (Ti, $T_C\simeq420$ mK). 
Despite the $\Phi_{ex}$ shift of $I_S^+$ and $I_S^-$ in these structures in well known \cite{Barone1982,Clarke2004,Duzer1999}, our SD is expected to be more attractive for applications than hybrid superconductor/semiconductor structures \cite{Ando2020,Miyasaka2021,Baumgartner2022,Baumgartner202b2,Pal2021,Gupta2022,Turini2022,Wu2022,Shin2021,Lin2021,Bauriedl2021,Merida2021}, thanks to the higher rectification efficiency, the absence of interfaces between different materials, and the ease of the fabrication protocol. 
Yet, although the only other fully-metallic supercurent diode developed so far shows larger values of $\eta$ \cite{Hou2022}, our SD has the additional advantage to be controlled  both with  a $\Phi_{ex}$ and $V_g$. 
In particular, a gate-controlled device guarantees, in principle, a higher switching speed than a magnetic field-tunable system. 
Finally, our SD shares materials composition and structure with widespread SCE architectures \cite{Duzer1999} such as cryotrons \cite{Buck1956}, nanocryotrons \cite{McCaughan2014}, RSFQs \cite{Likharev1991}, all-metallic gated devices\cite{DeSimoni2018,Paolucci2019,Ritter2021}, and memories \cite{Chen1992,Ligato2021,Murphy2017,Vernik2013}. Therefore, our flux- and gate-controlled SD geometry could find a number of different applications in superconducting electronics for the realization of high-speed and low-dissipation supercomputers.

\begin{acknowledgments}
The authors wish to acknowledge the EU’s Horizon 2020 research and innovation program under Grant Agreement No. 800923 (SUPERTED) and No. 964398
(SUPERGATE) for partial financial support.
\end{acknowledgments}

\section*{Data Availability Statement}
The data that support the findings of this study are available from the corresponding author upon reasonable request.

\providecommand{\noopsort}[1]{}\providecommand{\singleletter}[1]{#1}%

\end{document}